\documentclass[manuscript]{aastex6}

\usepackage{color}
\newcommand{\myemail}{takashi.tsukagoshi.sci@vc.ibaraki.ac.jp}

\begin{document}


\title{A Gap with a Deficit of Large Grains in the protoplanetary disk around TW Hya}

\author{
Takashi Tsukagoshi\altaffilmark{1},
Hideko Nomura\altaffilmark{2},
Takayuki Muto\altaffilmark{3},
Ryohei Kawabe\altaffilmark{4},
Daiki Ishimoto\altaffilmark{2,5},
Kazuhiro D. Kanagawa\altaffilmark{6},
Satoshi Okuzumi\altaffilmark{2},
Shigeru Ida\altaffilmark{7},
Catherine Walsh\altaffilmark{8},
T.J. Millar\altaffilmark{9}}

\altaffiltext{1}{College of Science, Ibaraki University, Bunkyo 2-1-1, Mito, Ibaraki, 310-8512, Japan; \myemail}
\altaffiltext{2}{Department of Earth and Planetary Sciences, Tokyo Institute of Technology, 2-12-1 Ookayama, Meguro, Tokyo, 152-8551, Japan}
\altaffiltext{3}{Division of Liberal Arts, Kogakuin University, 1-24-2 Nishi-Shinjuku, Shinjuku-ku, Tokyo, 163-8677, Japan}
\altaffiltext{4}{National Astronomical Observatory of Japan, 2-21-1 Osawa, Mitaka, Tokyo 181-8588, Japan}
\altaffiltext{5}{Department of Astronomy, Graduate School of Science, Kyoto University, Kitashirakawa-Oiwake-cho, Sakyo, Kyoto, 606-8502, Japan}
\altaffiltext{6}{Institute of Physics and CASA$^{*}$, Faculty of Mathematics and Physics, University of Szczecin, Wielkopolska 15, 70-451 Szczecin, Poland}
\altaffiltext{7}{Earth-Life Science Institute, Tokyo Institute of Technology, 2-12-1 Ookayama, Meguro, Tokyo 152-8550, Japan}
\altaffiltext{8}{Leiden Observatory, Leiden University, P. O. Box 9513, 2300 RA Leiden, The Netherlands}
\altaffiltext{9}{Astrophysics Research Centre, School of Mathematics and Physics, Queen's University Belfast, University Road, Belfast BT7 1NN, UK}

\begin{abstract}
We report $\sim$3 au resolution imaging observations of the protoplanetary disk around TW Hya at 145 and 233 GHz with the Atacama Large Millimeter/Submillimeter Array. Our observations revealed two deep gaps ($\sim$25--50 \%) at 22 and 37~au and shallower gaps (a few \%) at 6, 28, and 44~au, as recently reported by \citet{bib:andrews2016}.
The central hole with a radius of $\sim3$~au was also marginally resolved.
The most remarkable finding is that the spectral index $\alpha (R)$ between bands 4 and 6 peaks at the 22~au gap. 
The derived power-law index of the dust opacity $\beta (R)$ is $\sim1.7$ at the 22~au gap and decreases toward the disk center to $\sim0$.
The most prominent gap at 22~au could be caused by the gravitational interaction between the disk and an unseen planet with a mass of $\lesssim$1.5 $M_\mathrm{Neptune}$ although other origins may be possible.
The planet-induced gap is supported by the fact that $\beta (R)$ is enhanced at the 22~au gap, indicating a deficit of $\sim$mm-sized grains within the gap due to dust filtration by a planet.
\end{abstract}




\section{Introduction}
Protoplanetary disks are the birthplaces of planets.
The complex structures of protoplanetary disks such as spiral arms, inner holes, and gap and ring, recently reported by high-resolution infrared observations \citep[e.g.,][and references therein]{bib:espaillat2014}, are believed to be potential evidence of unseen planets in the disk.
Most recently, high-resolution observations with ALMA have found multiple gaps and rings in a disk even at submillimeter wavelengths \citep{bib:alma2015}.
Since submillimeter emission better traces the midplane density structures than infrared, the gaps and rings are thought to be direct evidence of the absence and enhancement of disk material, and therefore related to the planet formation process.
The origin of multiple gaps and rings is still under debate: several theoretical studies predict a formation scenario due to material clearance by planets \citep{bib:tamayo2015,bib:dong2015,bib:dipierro2015,bib:jin2016,bib:kanagawa2015,bib:kanagawa2016}, growth and destruction of icy dust aggregates near the snow lines of major volatiles \citep{bib:zhang2015,bib:okuzumi2016}, baroclinic instability triggered by dust settling \citep{bib:loren-aguilar2015} or secular gravitational instability \citep{bib:youdin2011,bib:takahashi2014}.\par

TW Hya is a 0.8 $M_\sun$ T Tauri star surrounded by a disk at a distance of $\sim$54 pc \citep[e.g.,][]{bib:andrews2012}.
Since the disk is almost face-on with an inclination angle of $7\degr$ \citep{bib:qi2004}, TW Hya is one of the best astronomical laboratories to investigate the radial structure of protoplanetary disks.
The disk mass has been measured to be $>0.05$ $M_\odot$ from HD line observations by the {\it Herschel} Space Observatory, indicating that it is massive enough to form a planetary system \citep{bib:bergin2013}.
Recently, a gap in the dust emission has been found at 20--30 au by submillimeter and near infrared observations \citep{bib:akiyama2015,bib:rapson2015,bib:debes2016,bib:nomura2016,bib:zhang2016}, which is possibly associated with the CO snow line \citep{bib:qi2013}.
Most recently, \citet{bib:andrews2016} reported the existence of multiple, axisymmetric gaps at 1, 22, 37, and 43 au at a spatial resolution of $\sim$1~au.
The depth and width of the submillimeter gap at 20--30 au are consistent with clearing by a super-Neptune mass planet \citep{bib:nomura2016}.
However, additional information on the dust size distribution with comparable spatial resolution is required to address the physical structure of the gap.
In this paper, we report multi-frequency observations of the disk around TW Hya with ALMA to probe the detailed disk structure and the change of dust spectral index across the dust gaps and rings at a spatial resolution of $\sim$3 au.

\section{Observations}
High-resolution continuum observations at bands 4 and 6 (145 and 233 GHz) with ALMA were carried out on 2015 December 1 and 2 (2015.A.00005.S).
In the observation period, 36 of the 12-m antennas were operational and the antenna configuration was in transition from C36-7 to C36-1, resulting in maximum baselines of 6.5 and 10.4 km for band 4 and 6, respectively.
We employed the Time Division Mode of the correlator, which is optimized for continuum observations.
The correlator was configured to detect dual polarizations in 4 spectral windows with a bandwidth of 1.875 GHz each, resulting in a total bandwidth of 7.5 GHz for each observed band.
The amplitude and phase were calibrated by observations of J1103-3251, and J1037-2934 was used for absolute flux calibration.
The observed passbands were calibrated by 5 min observations of J1037-2934 and J1107-4449 for band 4 and 6, respectively.\par

The visibility data were reduced and calibrated using the Common Astronomical Software Application (CASA) package, version 4.5.0.
After flagging bad data and applying the calibrations for bandpass, complex gain, and flux scaling, the corrected visibilities were imaged by the CLEAN algorithm.
The visibilities at band 6 with uv lengths $>$3000 k$\lambda$ were flagged out because of significant phase noise.
The uv sampling for baseline $\lesssim$400~m was particularly sparse along the north-south direction (i.e., v-axis of the uv coverage), which corresponds to $\lesssim180$ and $\lesssim300$~k$\lambda$ for band 4 and 6, respectively.
We have combined band 6 archival data (2012.1.00422.S), in which the maximum baseline is $\sim$500 k$\lambda$, with our band 6 data after applying a phase shift to account for proper motion and different input phase centers.
There were no available short-baseline data at band 4, hence only the long-baseline data were used for imaging.\par

To improve the image fidelity, we performed the iterative self-calibration imaging for each band data using the initial CLEAN image as a model image.
The interval time to solve the complex gain was varied from 600 to 90 sec for band 4 and from 1200 to 240 sec for band 6.
The resultant images after self-calibration were made by adopting briggs weighting of robust parameters 0.5 and 1.0 for band 4 and 6, respectively.
We also employed the multiscale clean with scale parameters of [0, 100, 300] and [0, 50, 150] mas for band 4 and 6, respectively, for better reconstruction of extended emission.
The spatial resolutions of the final images are 88.1$\times$62.1 mas with a position angle (PA) of $57\fdg8$ and 75.4$\times$55.2 mas with a PA of $38\fdg0$ for band 4 and 6, respectively.
The noise levels of the band 4 and 6 images are 12.4 and 28.7 $\mu$Jy beam$^{-1}$, respectively.

To deduce the spectral index between the bands 4 and 6 and to obtain a combined image around the center frequency (190 GHz), we also used the multi-frequency synthesis (MFS) method using all of the corrected visibilities after the iterative self-calibration imaging \citep[{\it nterm=2} in CASA {\it CLEAN} task; see][for the MFS method]{bib:rau2011}.
Briggs weighting with robust=0.0 was employed for the deconvolution and we also employed the multiscale option with scale parameters of 0, 60, and 180 mas.
Using the MFS method, we obtain the combined image and the map of the spectral index at 190 GHz.
The combined image achieves a better fidelity than the individual images since the observed data are combined to fill the gap in each other's uv coverage.
The achieved spatial resolution of the combined image is $72.7\times47.8$ mas, with a PA of $52\fdg9$, corresponding to $3.9\times2.9$ au.
The noise level is 15.9 $\mu$Jy beam$^{-1}$.



\section{Results}
\begin{figure*}\epsscale{1.0}
	\plotone{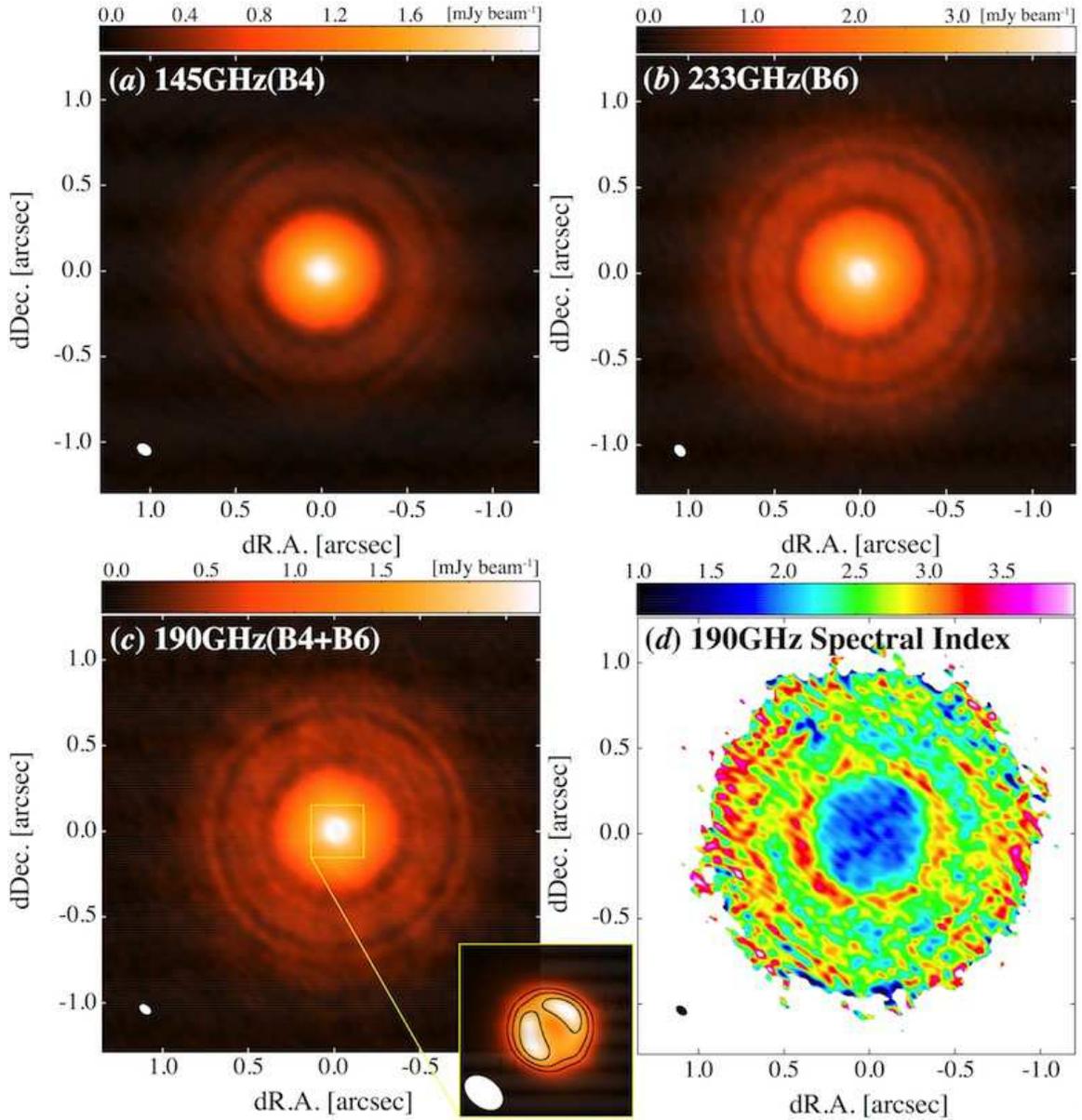}
	\caption{(a) and (b) show ALMA continuum images at 145 GHz (band 4) and 233 GHz (band 6), respectively. The ellipse at the bottom-left corner in each panel shows the synthesized beam. (c) shows the combined image of bands 4 and 6 with the MFS method. The inset indicates a close-up view ($0\farcs3\times0\farcs3$) for emphasis of the central structure. The contour indicates 130, 140, and 150$\sigma$. (d) shows the spectral index map derived from the MFS method.}\label{fig:obs}
\end{figure*}

Figures \ref{fig:obs}(a) and (b) show the constructed continuum maps at band 4 and 6, respectively.
Both images show circular multiple gaps and rings even though the resolution of the band 4 image is $\sim$1.3 times larger than that of band 6.
The total flux densities are 152.0$\pm$0.3 and 558.3$\pm$0.7 Jy for band 4 and 6, respectively.

Figure \ref{fig:obs}(c) shows the combined image of the band 4 and 6 data with the MFS method (hereafter MFS image).
The MFS image shows circular symmetric multiple gaps and rings.
In addition, we have resolved an inner hole with radius $\sim3$ au as predicted from an earlier analysis of the spectral energy distribution \citep{bib:calvet2002,bib:menu2014}.
This corresponds to the drop in the brightness temperature of dust continuum map recently found by \citet{bib:andrews2016}.
The total flux density integrated over the region with SN$>$10 is 360.3$\pm$0.5~mJy at 190~GHz (SN$\sim150$), which agrees well with the previous estimation at submillimeter wavelengths \citep{bib:qi2004,bib:andrews2012}.
There is no appreciable deviation from circular symmetry in the gaps, rings and spectral index $\alpha$.
Ellipsoid fittings of gaps and rings show the deviation between major and minor axes is within the errors ($\lesssim5\%$).\par

\begin{figure}\epsscale{0.8}
	\plotone{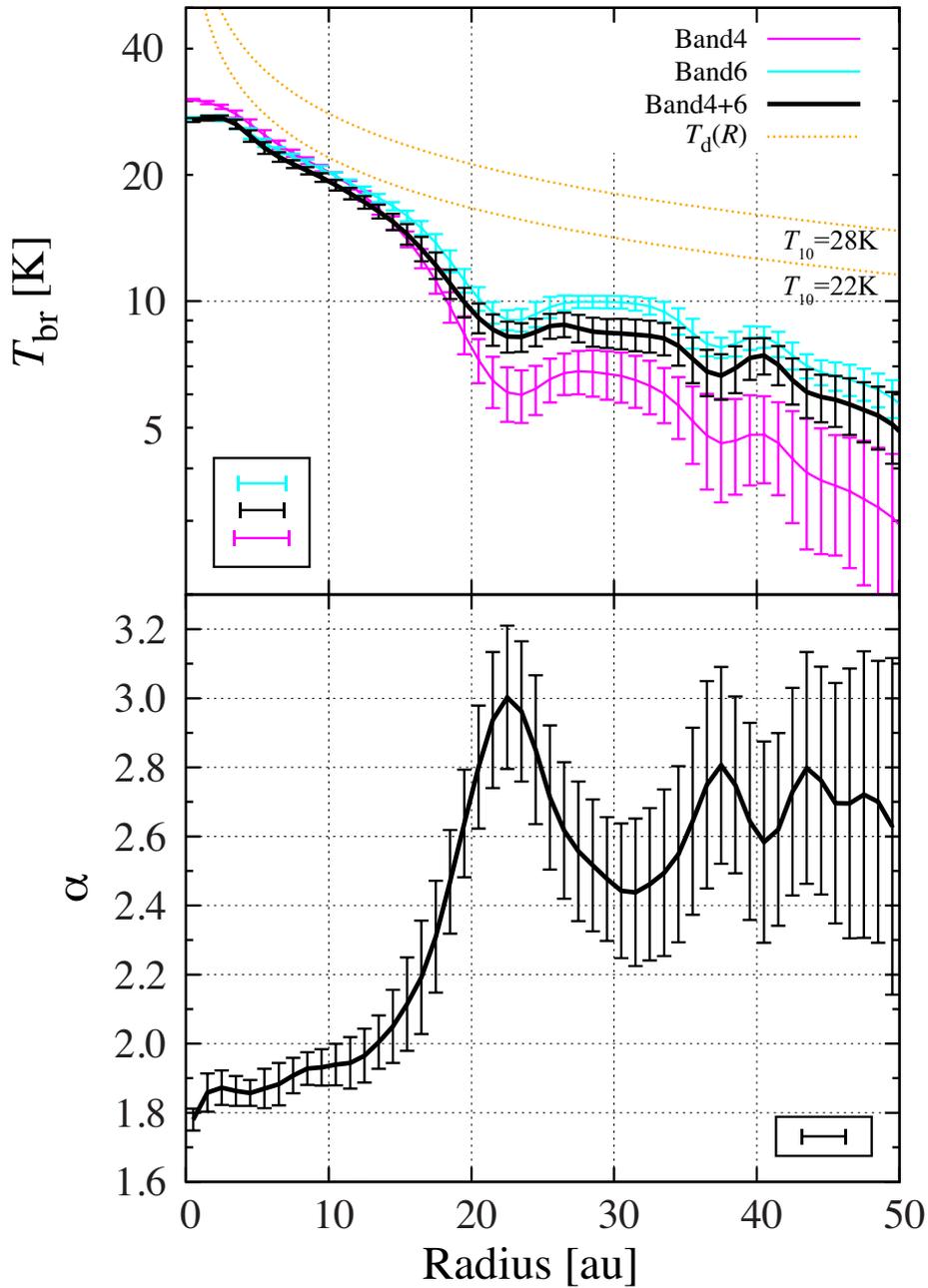}
	\caption{(top) Radial profiles of the brightness temperature averaged over full azimuthal angle. The black line indicates the MFS combined image and the lines in magenta and cyan show the bands 4 and 6 data, respectively. The bars at the bottom-left corner show the spatial resolutions. The error bar is determined from the standard deviation through the azimuthal averaging. The orange dashed lines indicate the assumed temperature profile of the dust disk when $T_{10}=$22 and 28~K with $q=0.4$. (bottom) Radial profile of the spectral index $\alpha$. The resolution is shown at the bottom-right corner in the panel. The error bar follows the same manner as the top panel.}\label{fig:radp}
\end{figure}

To confirm the gap structures, we plot the deprojected radial profile of the continuum emission in the top panel in figure \ref{fig:radp}.
The flux density is converted to the brightness temperature using the Planck function.
There are two prominent gaps at 22 and 37~au, and relatively weak decrements are also seen at 6, 28, and 44~au.
These observed features agree with those found by recent high-resolution ($\sim$1 au) observations at band 7 \citep{bib:andrews2016}.
The full width at half maximum and the relative depth are roughly 7~au and 50 \% for the 22~au gap if the background with a power-law form is assumed, and 3~au and $\sim20$ \% for the 37~au gap.
The depths are deeper than that of the gaps at band 7 \citep{bib:andrews2016}, while the widths are comparable.
Both the emission at band 4 and 6 show comparable brightness temperature inside $R\sim15$ au, the value of which is consistent with that of the band 7 emission \citep{bib:andrews2016}.
This result indicates that the disk is (at least moderately) optically thick in this region.\par

Figure \ref{fig:obs}(d) shows the spatial variation of the spectral index $\alpha$ (see eq. (\ref{eq:alpha}) for its definition).
The distribution seems to be axisymmetric, and therefore we make the radial profile of $\alpha$ averaged over the full azimuth angle as shown in figure \ref{fig:radp}.
The spectral index $\alpha$ radially decreases approaching the disk center.
There is a prominent peak around 22~au with $\alpha\sim3.0$, which coincides well with the position of the gap.
The enhancement in $\alpha$ is possible evidence of large grain deficit since $\alpha$ is related to the power-law index of the dust mass opacity if the emission is optically thin.
The rapid decrease of $\alpha$ inside the 22~au peak is partly due to increase of optical depth and partly due to decrease of the power-law index of $\beta$, namely, the existence of larger dust grains near the central star (see the next section).
There seems to be two weak ($<10$ \%) bumps at 37 and 44 au which are coincident with the locations of gaps in the intensity profile as for the 22~au gap, implying that there is a correlation between the surface brightness and $\alpha$. 

The error bars in figure \ref{fig:radp} are determined from the standard deviation determined by the azimuthal averaging.
This is a conservative way of determining the error because it is the most dominant source of the deviation at $>$5~au.
In fact, the uncertainty map for $\alpha$ produced by CASA shows an error is lower than the standard deviation.
The uncertainty in the absolute flux density does not affect the shape of the $\alpha$ profile, but the absolute scale of $\alpha$.
If the accuracy of the absolute flux scale is assumed to be $\sim10$\%, the $\alpha$ scale would have an associated error of $\Delta \alpha \sim 0.4$.
Therefore the weighted mean value is estimated to be $<\alpha>=2.42\pm0.42$, which agrees well with previous measurements for the entire disk \citep{bib:pinilla2014,bib:menu2014}.\par

\section{Discussion}
\subsection{Radial Profiles of Dust Optical Depth and Opacity $\beta$}
\begin{figure}\epsscale{1.0}
	\plotone{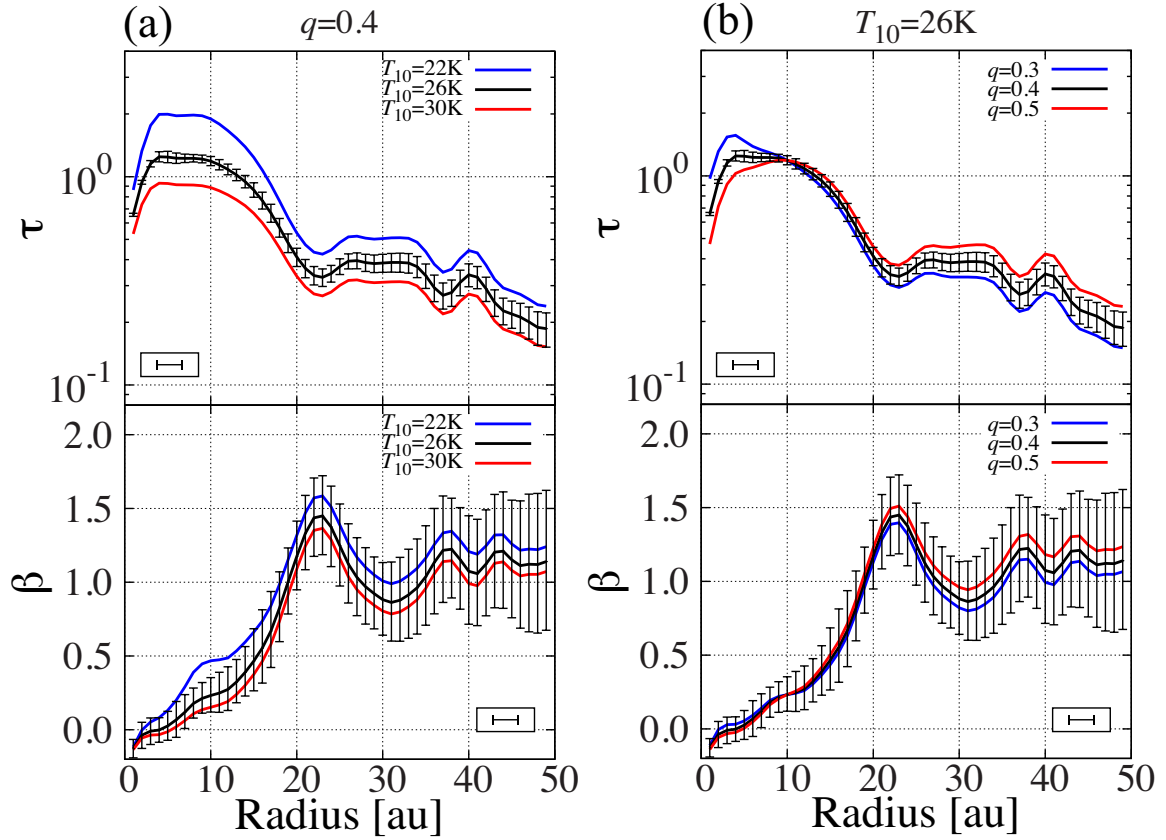}
	\caption{(a) Radial profile of the optical depth at 190 GHz (top) and $\beta$ (bottom). The cases for $T_{10}=$22, 26, and 30 K when $q$ is fixed to 0.4 are shown in blue, black, and red lines, respectively. The error bar is shown for the case of $T_{10}=$26~K representatively. The resolution is shown at the bottom-left (top) or bottom-right (bottom) corner in the panel. (b) Same as (a), but for checking the dependence on $q$ from 0.3 to 0.5 when $T_{10}$ is fixed to 26~K are shown in blue, black, and red lines, respectively. The error bar is shown for the $q=0.4 $ case representatively.}\label{fig:tau-beta}
\end{figure}

The intensity $I_{\nu}(R)$ and the spectral index $\alpha(R)$ are related to the dust temperature $T_d(R)$, the optical depth $\tau_{\nu}(R)$, and the dust opacity index $\beta(R)$ by
\begin{equation}\label{eq:intensity}
 I_\nu (R) = B_\nu(T_d(R)) \left( 1-\exp \left[ - \tau_\nu \right] \right)
\end{equation}
and 
\begin{equation}\label{eq:alpha}
 \alpha(R) \equiv \frac{d \log(I_{\nu})}{d \log \nu} 
  = 3 - \frac{h\nu}{k_BT_d(R)} \frac{e^{h\nu/k_BT_d(R)}}{e^{h\nu/k_B T_d(R)} - 1}
    + \beta(R) \frac{\tau_{\nu}(R)}{e^{\tau_{\nu}(R)}-1}.
\end{equation}
Here, $B_{\nu}(T)$ is the Planck function, $h$ is Planck's constant, $c$ is the speed of light and $k_B$ is Boltzmann's constant.
The optical depth is assumed to have the form $\tau_\nu(R)=\tau_\mathrm{190GHz}(R)(\nu/\mathrm{190GHz})^\beta$.
There are three unknown variables in equations (\ref{eq:intensity}) and (\ref{eq:alpha}), which are $T_d(R)$, $\tau_\mathrm{190GHz}(R)$, and $\beta (R)$.
If we assume one of them, we are able to derive the rest of them by using the observation data of $I_\nu (R)$ and $\alpha (R)$.
Here, we assume that $T_d (R)$ is given by $T_d (R) = T_{10} (R/10~\mathrm{AU})^{-q}$.
We vary $T_{10}$ from 22 to 30~K and $q$ from 0.3 to 0.5 to see how the temperature affects the derived physical quantities.
This assumption is based on our fitting to the temperature profile at the disk midplane in \citet{bib:andrews2012} and \citet{bib:andrews2016}.
The temperature profile models are chosen so that the observed brightness temperature does not exceed the kinetic temperature.
The assumed temperature profiles, however, have no great impact on the following conclusions as shown in figure \ref{fig:tau-beta}.

Figure \ref{fig:tau-beta} shows the radial profiles of $\tau_\mathrm{190GHz} (R)$ and $\beta (R)$.
The errors are estimated in a conservative way in which the combination between the maximum and minimum values of the error bars in the intensity and $\alpha (R)$ profiles is used for determining the maximum range of the error.
The disk is optically thin at $R>15$~au in all the cases and marginally optically thick at $R<$15~au.
This is in contrast with HL Tau \citep{bib:alma2015,bib:pinte2016}, where an optically thick region extends out to $R\lesssim40$~au.
We see a prominent drop in the optical depth at $R<5$~au, which likely corresponds to the inner hole derived from the SED \citep{bib:calvet2002} and to the drop in the brightness temperature of the dust continuum map recently found by \citet{bib:andrews2016}.
The optical depth profiles have two dips at $R \sim22$~au and  $\sim37$~au.
Note that although $\beta$ can not be determined where the optical thickness is considerably high, the $\beta$ profile at $<$15~au is still accessible because $\tau$ is of order unity.\par

Overall, $\beta (R)$ increases from $\sim$0 to $\sim$1.7 with when moving from the disk center to $\sim$20~au, where the disk is marginally optically thick.
This implies that sufficient large dust grains ($\geq$10~mm) exist at 5--10~au.
Radially increasing profiles of $\beta (R)$ are also seen in other T Tauri disks \citep[e.g.,][]{bib:perez2012}, and compact distribution of the largest grains is suggested in the TW Hya disk, too \citep{bib:menu2014}.\par

One of the most remarkable features of the $\beta (R)$ profile is the peak at $\sim$22~au, which corresponds to the location of the gap in the surface brightness profile.
This indicates that large dust grains are less abundant within the gap compared to other locations in the disk.
We also tentatively see the increase in $\beta (R)$ near the 37~au gap (and perhaps also near the 44~au gap), but further observations with better sensitivity is needed to confirm this.

At $R<$15~au where $\tau(R) \sim 1$, $\beta(R)$ is derived to be ranging from 0.0 to 0.5, and according to the theoretical calculation of dust mass opacity \citep{bib:draine2006}, small $\beta$ value suggests that the power-law index of dust size distribution is very small and and the maximum dust size is large ($>$a few cm).
Also, the result suggests that  the column density would be at least an order of magnitude higher than that at the 22~au gap, giving us the column density with an power-law  index of $<-2$.
The steep profile is consistent with the previous measurement that large grains should be concentrated towards the inner disk region to reproduce the 9 mm emission \citep{bib:menu2014}.\par

The discussions of $\tau_\mathrm{190GHz} (R)$ and $\beta (R)$ presented here are based on the assumptions of smooth temperature profiles.
Observations at additional bands (preferably at lower frequencies) may further constrain $T_d (R)$, $\tau_\nu (R)$ and $\beta (R)$ simultaneously.
We note that our results are roughly consistent with the Band 7 observations by \citet{bib:andrews2016}.

\subsection{Origin of the Gaps}
The enhancement of $\beta (R)$ indicates a deficit of large (mm-sized) grains at the gap position.
These facts support the scenario that the gap is caused by planet-disk interaction because it is consistent with the picture of dust filtration and trapping due to a planet \citep{bib:zhu2012}.
Using the relationship that connects the gap shape with the planet mass \citep{bib:kanagawa2015,bib:kanagawa2016}, a planet with 1.5 $M_\mathrm{Neptune}$ may be responsible for the gap, assuming a viscosity parameter $\alpha=10^{-3}$ and a disk aspect ratio of 0.05 (consistent with the assumption of $T_{10}=22~\mathrm{K}$).
We note that similar values are derived from both gap width and depth.
This planet mass should be considered as the upper limit since the formula by \citet{bib:kanagawa2015,bib:kanagawa2016} is for the gas gap and the actual dust gap may be wider and deeper than the gas gap due to dust filtration \citep{bib:zhu2012}.


Alternatively, the multiple ring structures might be related to the snow lines of major volatiles \citep{bib:zhang2015,bib:okuzumi2016}.
TW Hya is suggested to have a CO snow line at $\sim$30~au \citep{bib:qi2013,bib:schwarz2016}, and our observations identify a bright dust ring near this snow line.
This is consistent with the dust ring formation scenario by \citet{bib:okuzumi2016}, in which icy dust aggregates experience sintering, disrupt, and pileup near major snow lines.
As noted by \citet{bib:andrews2016}, the 40~au bright ring might correspond to the snow line of N$_2$, which has a sublimation temperature slightly lower than that of CO.
However, the model of \citet{bib:okuzumi2016} does not predict a strong radial variation of $\beta (R)$, thus not explaining the enhancement of $\beta (R)$ we found near the 20~au dark ring.\par

The multiple gaps with intervals of 5--10~au beyond the 22~au gap (22, 28, 37, and 44~au) may be reminiscent of dynamical instabilities within the disk such as zonal flow patterns driven by MHD turbulence \citep{bib:johansen2009}, baroclinic instability driven by dust settling \citep{bib:loren-aguilar2015}, and/or the secular gravitational instability \citep{bib:youdin2011,bib:takahashi2014}.
Different dynamical processes act under different physical conditions and therefore, better constraints on the dust disk physical structure based on high resolution observations at other bands \citep[e.g.,][]{bib:andrews2016} and constraints of the density and temperature structures of gas component are essential in determining the origin of such structures.

\acknowledgments
We thank the referee for many useful comments and critiques which helped improve the contents of this paper.
This paper makes use of the following ALMA data: ADS/JAO.ALMA\#2015.A.00005.S and ADS/JAO.ALMA\#2012.1.00422.S. ALMA is a partnership of ESO (representing its member states), NSF (USA) and NINS (Japan), together with NRC (Canada), NSC and ASIAA (Taiwan), and KASI (Republic of Korea), in cooperation with the Republic of Chile. The Joint ALMA Observatory is operated by ESO, AUI/NRAO and NAOJ. A part of data analysis was carried out on common use data analysis computer system at the Astronomy Data Center of NAOJ. This work is partially supported by JSPS KAKENHI grant numbers 24103504 (TT), 23103005 and 25400229 (HN), 26800106 and 15H02074 (TM), and 16K17661 (SO). KDK was supported by Polish National Science Centre MAESTRO grant DEC- 2012/06/A/ST9/00276. Astrophysics at QUB is supported by a grant from the STFC.




\end{document}